\documentclass[aps,prl,twocolumn,showpacs,groupedaddress,amsmath,amssymb]{revtex4}
\usepackage{graphicx}
\usepackage{bm}
\usepackage{doi}
\usepackage{float}
\begin{document}
\draft
\def\ds{\displaystyle}
\title{  Robust Exceptional Points in Disordered Systems  }
\author{Cem Yuce$^1$}
\email{cyuce@anadolu.edu.tr}
\author{Hamidreza Ramezani$^2$}
\affiliation{$^1$Department of Physics, Eskisehir Technical University, Turkey\\
$^2$Department of Physics and Astronomy, University of Texas Rio Grande Valley, Brownsville, TX 78520, USA}

\date{\today}
\begin{abstract}

We construct a theory to introduce the concept of topologically robust exceptional points (EP). Starting from an ordered system with $N$ elements, we find the necessary condition to have the highest order exceptional point, namely $N^\text{th}$ order EP. Using symmetry considerations, we show an EP associated with an order system is very sensitive to the disorder. Specifically, if the EP associated with the ordered system occurs at the fixed degree of non-Hermiticity $\gamma_{EP}$, the disordered system will not have EP at the same $\gamma_{PT}$ which puts an obstacle in front of the observation and applications of EPs. To overcome this challenge, by incorporating an asymmetric coupling we propose a disordered system that has a robust EP which is extended all over the space. While our approach can be easily realized in electronic circuits and acoustics, we propose a simple experimentally feasible photonic system to realize our robust EP. Our results will open a new direction to search for topologically robust extended states (as opposed to topological localized states) and find considerable applications in direct observation of EPs, realizing topological sensors and designing robust devices for metrology.  

 \end{abstract}
\maketitle

{\it Introduction}-- Non-Hermitian systems present new functionalities and features that are inaccessible in Hermitian systems. For instance the existence of exceptional points (EP) is a unique feature of the non-Hermitian systems that resulted to many fascinating features including unidirectional invisibility \cite{unitran2,unitran3}, unidirectional lasing \cite{unitran1}, lasing and anti-lasing in a cavity \cite{unitran4}, enhanced optical sensitivity \cite{sense1,sense2} and stopping of light \cite{listop}. 
An EP is a topological singular point in the parameter space of a non-Hermitian Hamiltonian and occurs when at least two eigenvalues and the corresponding eigenvectors of the Hamiltonian coalesce \cite{EP1,EP2,EP3} meaning that the Hamiltonian becomes defective. It can be shown that during the encircling of an EP using a closed loop, the eigenstates of the corresponding Hamiltonian become swapped \cite{dynexcep1,dynexcep2}.
The topological nature of the EP is a consequence of the fact that swapping of the eigenstates is independent of the loop and it occurs even if we deform the loop.  In particular, in parity-time symmetric systems with carefully distributed balanced amplification and absorption mechanisms, an EP determines a phase transition from a real spectrum to a complex spectrum. 
Usually, EPs are associated with the states that are distributed all over the space and thus EPs are sensitive to the unintentional and yet unavoidable changes in the parameter space. As a result, direct observation of EPs is a difficult task. Despite the challenges, their influence on the dynamical properties of the system has been experimentally realized \cite{EPdeney1,EPdeney2,EPdeney3,EPdeney4}. 

Due to their topological nature, EPs are believed to play an essential role in the theory of topological insulators in non-Hermitian systems. The extension of topological phases to non-Hermitian systems is a relatively new field of study and has recently attracted a great deal of attention. This includes theoretical studies of topological edge states in $1$-D non-Hermitian systems \cite{nonh2,PTop3,ekl56,PTop4}. These theoretical works are followed by the first experimental realization of dynamically and topologically stable zero energy state in $1$-D using waveguides \cite{sondeney1}. The main model studied in the context of non-Hermitian topological insulators is the $1$-D Su-Schrieffer-Heeger (SSH) model with gain and loss \cite{1d1,1d2,1d3}. The $1$-D study of topological phases has been extended to higher dimensions such as $2$-D photonic honeycomb lattice with gain and loss \cite{2d1,2d3} and a non-Hermitian extension of the Bernevig, Hughes, and Zhang (BHZ) model \cite{2d4}. A major conclusion of these works shows that non-Hermitian systems don't respect the standard bulk-boundary correspondence of topological insulating systems \cite{bulkboun01,bulkboun02,bulkboun03,bulkboun04,bulkboun05,bulkboun06,bulkboun07}. For example, topological transition point of the open periodic systems can be different in non-Hermitian systems in contrast to their corresponding Hermitian ones. Furthermore, standard formulation of topological numbers such as winding and Chern numbers don't perfectly work in non-Hermitian systems \cite{winding1,winding2,winding3}. Not only topological insulators but also other topological systems have been studied in the context of non-Hermitian systems. These are, for example, topological superconductors with gain and loss and Majorana modes \cite{majo1,majo2,majo3,majo4,majo5,majo6}, nodal surfaces formed by EPs \cite{nodalkun,nodalkun2} and Floquet topological phase in non-Hermitian systems \cite{floquet1,floquet2,floquet3,floquet4}.

As mentioned earlier, from the experimental point of view, accessing EPs is a difficult task. Unfortunately, unavoidable disorders in an experiment shift the EPs in parameter space drastically which makes it almost impossible to access an EP. In this paper, we provide a general theory for the creation of a robust EP. While our theory can be applied to different systems including acoustic, phononic, and electronic systems, we provide a simple model for the realization of such robust EPs in a non-Hermitian photonic system. 



{\it Formalism}-- In a Hermitian system, the spectrum is purely real and degenerate points in the spectrum, the so-called diabolical points, are associated with orthogonal and distinct eigenstates. In contrast in a non-Hermitian system, degenerate points or EPs in the spectrum have a degenerate eigenstate due to the coalescing of the corresponding eigenstates \cite{EP1}.
For instance, consider a finite system that is described by an $N{\times}N$ non-Hermitian Hamiltonian $\mathcal{H}_{N\times N}$. An $n$-th order exceptional point is formed by the simultaneous coalescence of $n$ eigenstates. It is obvious that an $N{\times}N$ non-Hermitian Hamiltonian can have at most $N$-th order exceptional point. 

{\it $N$-th order exceptional points}--
 Apart from fundamental curiosity higher order exceptional points are shown to be good nominees for sensing and metrology\cite{lyang1, mkhaje}. So it would be important to find the sufficient condition of realizing an $N$-th order EP. To answer this question without losing generality we search for Hamiltonians that have an $N$-th order exceptional point with zero energy eigenvalue. In other words we are looking for Hamiltonians with a spectrum given by $N$-th root of an expression that becomes zero at the EP. Let us consider the following non-Hermitian Hamiltonians with 
 $N=2,3$ and $4$:
  \begin{equation}\label{yudj2}
 \begin{split}
 \mathcal{H}_{2\times 2}=& i\gamma\sigma_z+\sqrt{2}\sigma_x\\
 \mathcal{H}_{3\times 3}= &\left(\begin{array}{ccccc}i\gamma &1& 0   \\ 1& 0 &1   \\0&1 &  -i\gamma  \end{array}\right)\\
 \mathcal{H}_{4\times 4}=&\left(\begin{array}{cc}i\gamma\sigma_z+\sigma_x &-i\sigma_y\\i\sigma_y&i\gamma\sigma_z+\sigma_x   \end{array}\right)
\end{split}
 \end{equation}
 where $\gamma$ is the degree of non-Hermiticity and $\sigma_i$ are the Pauli matrices. All the above Hamiltonians in Eq.(\ref{yudj2}) share a same feature namely all of them are {\textit{nilpotent}} of order $N$ for some value of $\gamma$. A finite-dimensional Hamiltonian $\mathcal{H}$ is a nilpotent operator of order $p$ if $\mathcal{H}^p $ is equal to the zero operator for some positive integer $p$, namely $\mathcal{H}^p=\mathbf{0}$. A nilpotent Hamiltonian $\mathcal{H}$ is always noninvertible, namely $Det(\mathcal{H})=0$, and all of its eigenvalues are equal to $0$. The later can be easily shown when we operate the Hamiltonian $\mathcal{H}$ on its nonzero eigenvector $|\lambda\rangle$ with eigenvalue $\lambda$ for $p$ times
 \begin{equation}\label{eq2}
\mathcal{H}^p|\lambda\rangle=\lambda^p|\lambda\rangle
 \end{equation} 
 However, the left hand side of Eq.(\ref{eq2}) is zero operator $\mathbf{0}$ with eigenvalue zero. Thus, the $p$-th root of $\lambda$ is zero. Consequently, if a non-Hermitian Hamiltonian is not nilpotent of order $p<N$, then the order of (zero energy) exceptional states are less than $N$. The most simple example of a nilpotent matrix is an upper triangular matrix, the so-called Jordan block matrix. 
Now let us go back to the Hamiltonians in Eq.(\ref{yudj2}). The first two Hamiltonians have open edges while the edges for the last system are connected with antisymmetric tunneling amplitude. We observe that $ \mathcal{H}_{2\times2}^2  = \mathcal{H}_{3\times3}^3  = \mathcal{H}_{4\times4}^3  =\mathbf{0}$ when $\gamma=\sqrt{2}$. Therefore, all these three Hamiltonians are nilpotent with zero energy exceptional eigenstate at $\gamma=\gamma_{EP}=\sqrt{2}$. Note that the exceptional point $\gamma_{EP}$ is also the transition point from real valued eigenvalues to complex ones. 

Our above discussion depicts that a $N{\times}N$ non-Hermitian Hamiltonian must be nilpotent for the existence of $N$-th order {\it zero} energy exceptional states. On the other hand if the $N$-th order energy exceptional state has a nonzero eigenvalue it does not necessarily need to be nilpotent. Any nonzero exceptional point can be formed by $\mathcal{H}^{\prime} =\mathcal{H}+\alpha I$, where the Hamiltonian $\mathcal{H}$ is nilpotent, $\alpha$ is an arbitrary complex valued constant and $I$ is the $N{\times}N$ identity matrix. This new Hamiltonian is not nilpotent since $Det\left(\mathcal{H}^{\prime}\right)\neq0$. One can easily see that $N$-th order exceptional eigenstate of $\mathcal{H}$ is the simultaneous eigenstate of $\mathcal{H} ^{\prime}$ with eigenvalue $\alpha$. Therefore, we conclude that nonzero energy exceptional states appear in a system that is not nilpotent. As an example, consider the following Hamiltonian with an asymmetric tunneling amplitude and an on-site potential difference between the sites: $\mathcal{H}=\sigma_z+i\sigma_y$. This Hamiltonian is nilpotent of order $2$ and hence it has a second order exceptional state with zero energy. Now, let us consider  $\mathcal{H}+\alpha\mathcal{I}$, which is not nilpotent.
It is easy to see that both Hamiltonians have a simultaneous exceptional eigenstate
$\frac{1}{\sqrt{2}} \left(\begin{array}{cc}-1 &1\end{array}\right)^T$. However, it is important to notice that such nonzero energy EP is a trivial shift of the levels and does not imply significant physics.

{\it Symmetry properties}-- So far we provided the condition for obtaining an $N$-th order exceptional point. To illustrate the concept of topologically protected EP, it is constructive to understand the relation between the symmetry properties of a non-Hermitian Hamiltonian $\mathcal{H}(\gamma)$ and the reality of its energy eigenvalues, assuming $\mathcal{H}(\gamma)$ is non-Hermitian Hamiltonian that can become nilpotent for some value of $\gamma=\gamma_{EP}$. More specifically, the exceptional point $\gamma_{EP}$ is a transition point from real-valued to complex-valued energy
eigenvalues for the nilpotent Hamiltonian $\mathcal{H}(\gamma_{EP})$. We stress that a nilpotent Hamiltonian is not necessarily a parity-time ($\mathcal{PT}$) symmetric operator and its energy eigenvalues are always zero. Specifically, $\mathcal{PT}$ symmetry is neither a necessary nor a sufficient condition for the existence of the nilpotent
Hamiltonian $\mathcal{H}(\gamma)$. In other words,
$\mathcal{PT}$ symmetry is not restored below or beyond the
EP of $\mathcal{H}(\gamma)$. 

Generalization of the above conclusion to higher dimension is straightforward, namely a non-Hermitian Hamiltonian $\mathcal{H}(\gamma_1,\gamma_2,...)$ with more than one parameter becomes nilpotent on a
exceptional curve, a exceptional surface, and a exceptional hyper-surface
for the $2$, $3$, and larger dimensional parameter spaces,
respectively. Obviously, $\mathcal{PT}$ symmetry condition is not
generally satisfied for this case, either. As a simple example,
consider a two-level Hamiltonian $\mathcal{H}(\gamma_1,\gamma_2)=i\sigma_z+\frac{\gamma_1+\gamma_2}{2}\sigma_x+\frac{i(\gamma_1-\gamma_2)}{2}$ with asymmetric
tunneling amplitudes $\gamma_{1,2}$ and a balanced gain and loss coefficient equal to one. This Hamiltonian
is parity-time $\mathcal{PT} =\sigma_y \mathcal{K}$ ($\mathcal{PT} =\sigma_x \mathcal{K}$) symmetry when
$\gamma_1=-\gamma_2$ ($\gamma_1=\gamma_2$), where
$\mathcal{K}$ is the complex conjugation operator. For any other values of $\gamma_{1,2}$, namely $\gamma_1\neq\mp\gamma_2$ this Hamiltonian
is no longer $\mathcal{PT}$ symmetric.
Despite the fact that $\mathcal{H}(\gamma_1,\gamma_2)$ might not be $\mathcal{PT}$ symmetric it admits real-valued eigenvalues for $\gamma_2\geq1/\gamma_1$ and complex eigenvalues for otherwise. Furthermore, the exceptional curve with zero energy is
determined by $\gamma_2=1/\gamma_1$ and the corresponding eigenstate $\frac{1}{\sqrt{1+\gamma_1^2}}
(\begin{array}{cc}i\gamma_1 &1\end{array})^T$. Note that the Hamiltonian $\mathcal{H}$ becomes nilpotent at $\gamma_2=\gamma_1$ and a $2^{\text{nd}}$ order EP appears.

\begin{figure}
	\includegraphics[width=1\linewidth, angle=0]{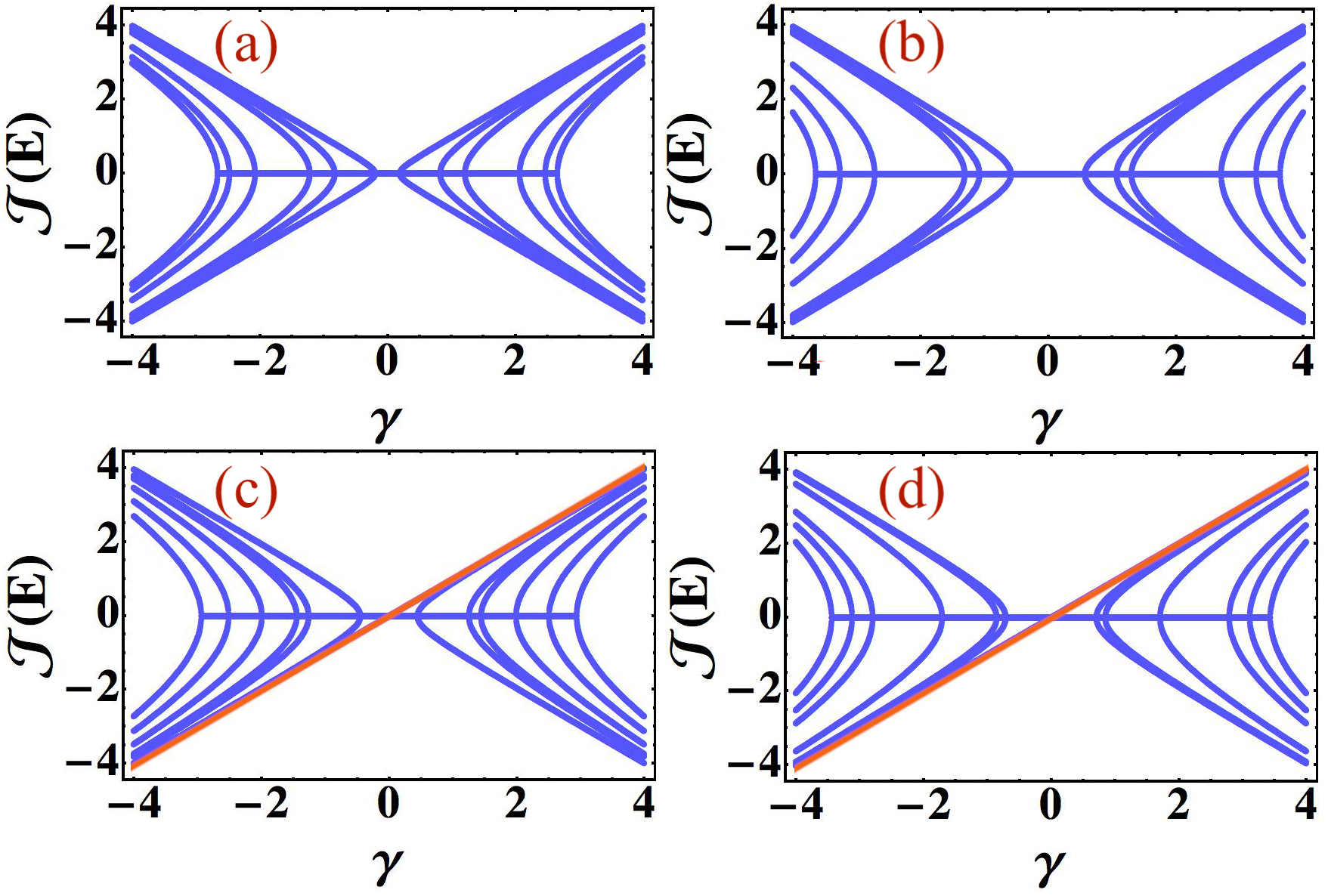}
	\caption{ Imaginary part of energy eigenvalues for $N=12$ (top panel) and $N=13$ (lower panel) for the Hamiltonian (\ref{plkjhmu}) with $t_{n,n\mp1}\in[0.5,2]$ (a,c) and  $t_{n,n\mp1}\in[0.5,3]$ (b,d). The eigenstates with energy eigenvalue $i\gamma$ for $N=13$ represented by the orange line is topologically stable. The branch points are associated with EP points and the determinant of the corresponding Hamiltonians are zero at these exceptional points.}
	\label{fig1}
\end{figure}
Another example of non-Hermitian systems with EPs and without $\mathcal{PT}$ symmetry is a non-Hermitian disordered system. In general a disordered system can't be described by a nilpotent Hamiltonian and hence they can't have $N$-th order EP.
However, second order EP can occur in a disordered system. To illustrate this statement, consider a disordered
tight binding lattice with alternating gain and loss
\begin{eqnarray}\label{plkjhmu}
\mathcal{H}&=& \sum_{n=1}^{N-1} t_n(|n\rangle \langle n+1|+|n+1\rangle \langle n|)+
\nonumber\\ &&\sum_{n=1}^{N} (-1)^ni\gamma |n\rangle \langle n|
\end{eqnarray}
where $t_n$ is the tunneling amplitude drawn randomly from a uniform box distribution and $\gamma$ is the non-Hermitian degree. In Fig.(\ref{fig1}) we are plotting the imaginary parts of the energy eigenvalues associated with the $\mathcal{H}$ in Eq.(\ref{plkjhmu}) for $N=12$ (top panel) and $N=13$ (lower panel) as a function of $\gamma$ for two different distributions $t_{n}\in[0.5,2]$ (left panel)
and $t_{n}\in[0.5,3]$ (right panel). We observe
that there exist $M\in\left[ 1,N\right]$ distinct $\gamma_{EP}$ values (branch points associated with the second order Eps). The number $M$ and the value of $\gamma_{EP}$ are sensitive to the disorder. Namely, every time that we set the tunneling amplitudes from a uniform distribution box and we vary the $\gamma$, EPs are found at different value with respect to another set of tunneling. However, as it can be seen from the lower panel of Fig.(\ref{fig1}), there is a distinctive eigenvalue
$E=i\gamma$ (highlighted with a orange line) for odd values of $N$.
This arises in the systems with $N \equiv2m-1$ (where $m\in \mathbb{N}$) since gain and loss are not balanced and they do not form a set of dimers with amplification and absorption. Furthermore, eigenenergy of this state is not affected by tunneling disorders, namely its exact energy eigenvalue $E=i\gamma$ does not disturb when we change the tunneling strength randomly. This is in contrast to other eigenenergies that are sensitive to tunneling disorder. Therefore, we can conclude that for odd values of $N$ this state is a
topological state. It is important to notice that this state is {\it not a localized state} on either side of the lattice and is extended through the whole lattice with zero amplitude on the even sites.

From the above discussion, it is clear that while second order EPs are not destroyed by the disorder, in general they are not robust against the tunneling
amplitude disorder. In other words the positions of them on the $\gamma$ axis, namely $\gamma_{EP}$ values, are changing with the disorder. The only robust point in the Fig.(\ref{fig1}) is associated with $\gamma=0$ point, which is not an EP as the Hamiltonian becomes Hermitian. Thus in general, exceptional points are not robust in the presence of disorder. However, this does not mean that one cannot find a robust EP. Below we propose a system that has a robust EP in the presence of the disorders.

\begin{figure}
\includegraphics[width=1\linewidth, angle=0]{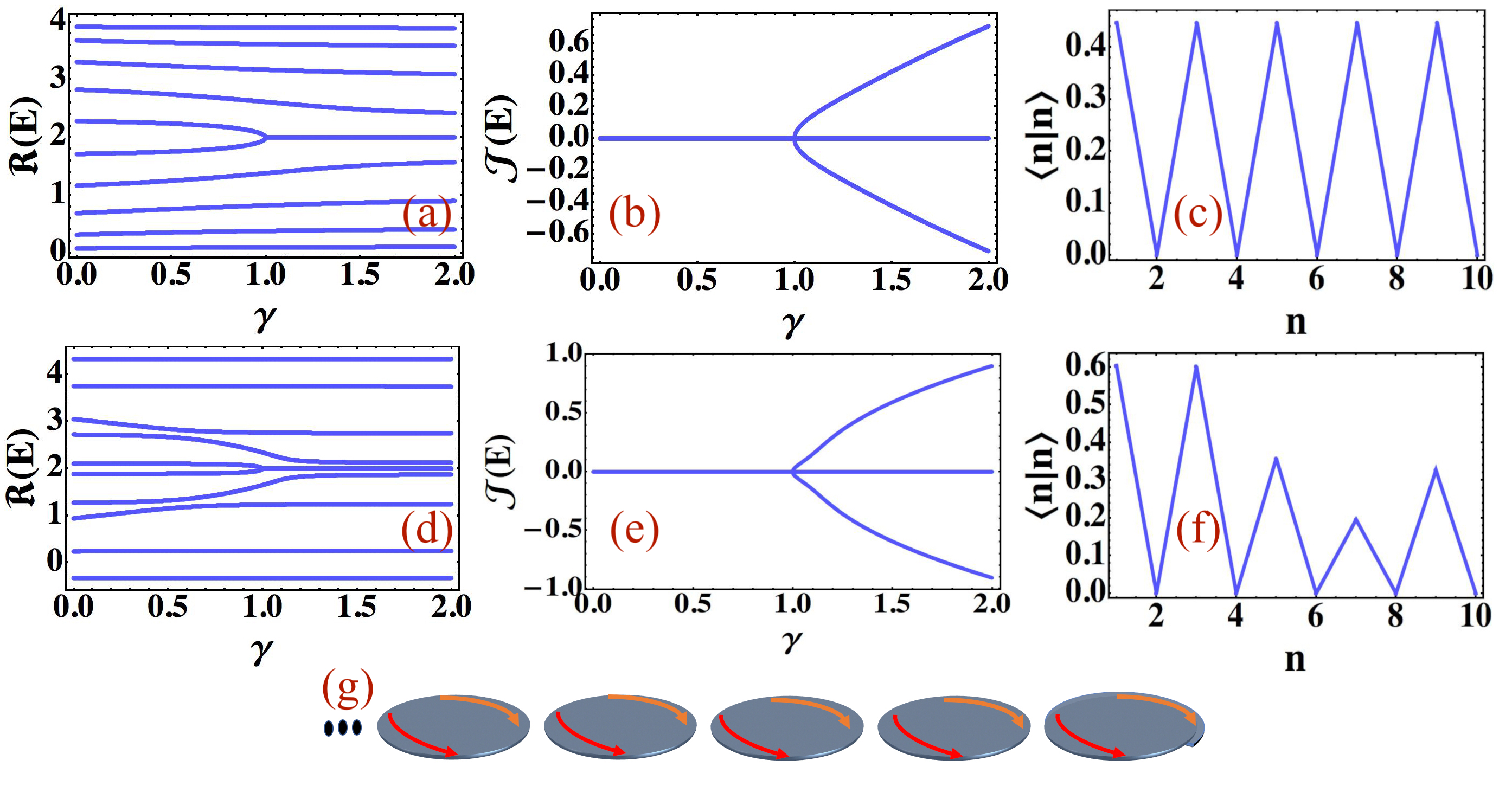}
\caption{ The real (a) and imaginary (b) parts of energy eigenvalues and the mode profile (c) associated with the EP of the Hamiltonian (\ref{ysonmu}) for $\beta=2$, $N=10$ and $t_n=1$. (d-f) similar to (a-c) in the presence of strong disordered with $t_{n}\in1+[-0.75,0.75]$ and $t_{N-1}=1$. The real and imaginary parts of energy eigenvalues are deformed in the presence of the disorder. However, the exceptional point at $\ds{\gamma=1}$ is robust against the disorder. (g) Schematic of a proposed structure with robust EP which is composed of a $N-1$ symmetric resonator chain  coupled from the right side to an asymmetric resonator with snail shell form. Each resonator support two modes, CW and CCW. Only in the last resonator on the right side the CW and CCW modes of the same resonator are coupled to each other in an asymmetric form.}
\label{fig2}
\end{figure}

{\it Robust exceptional points}-- Consider the following $1D$ non-Hermitian tight binding Hamiltonian with on-site potential $\beta$ and symmetric tunneling amplitude $t_n=1$ everywhere except at its right edge which has an asymmetric coupling.
\begin{eqnarray}\label{ysonmu}
\mathcal{H}&=& \sum_{n=1}^{N-1}t_n |n\rangle \langle n+1|+j_n|n+1\rangle\langle n|+\sum_{n=1}^{N} \beta |n\rangle \langle n|
\end{eqnarray}
where $t_n$ and $j_n$ are forward and backward tunneling which in general they can be complex (random) numbers. If $t_n=J_n^*$, then the system becomes Hermitian otherwise it is non-Hermitian. Let us now assume that the chain support symmetric couplings except at the last coupling, viz.
\begin{equation}
j_{n}=\begin{cases}
t_n&\text{if }  n\neq N-1\\
t_{n}-\gamma& \text{if } n=N-1
\end{cases}
\label{eq5}
\end{equation}
Where in Eq.(\ref{eq5}) the parameter $\gamma$ denotes the strength of the asymmetry in the last coupling. Without loss of generality we assume that $N$ is an even number. The Hamiltonian in Eq.(\ref{ysonmu}) becomes non-invertible when $\beta=0$. As depicted in the top panel of Fig.(\ref{fig2}), this system for $t_n=1$ has a second order EP which appears when $\gamma=1$. More specifically, the energy eigenvalues are real (complex) valued when $\gamma<1$ ($\gamma>1$). This EP is robust against tunneling amplitude disorder as long as $\gamma=1$ and $t_{N-1}=1$. 
To show the robustness of this EP we perform numerical computation on the disordered system with $\beta\neq 0$ (which makes $\mathcal{H}$ invertible) \cite{note} and $t_{n}\in1+[-0.75,0.75]$ while we keep $t_{N-1}=1$. An example of the real and imaginary part of the eigenvalues of this disordered system is shown in the lower panel of Fig. (\ref{fig2}). As can be seen, the energy eigenvalues are deformed due to the disorder but the EP at $\gamma=1$ remains the same, meaning it is robust against disorder. We stress that the eigenstate associated with the EP is an extended system as shown in Figs(\ref{fig2}c,f). 

The above intrinsic model can be realized in a simple passive photonic resonator chain. Especially, consider a chain of coupled-resonator optical waveguide which consists of identical symmetric coupled cavities with resonance frequency $\omega_0$ \cite{52,53,54}. Now let us couple this chain to a single cavity with resonance frequency $\omega_0$ that is no longer symmetric and has a {\it snail shell form} as sketched in Fig.(\ref{fig2}g) \cite{55}. Each resonator features a clockwise (CW) and a counterclockwise (CCW) mode. While the CW and CCW modes of a symmetric cavity are decoupled there is a coupling $t$ between the CW (CCW) of a cavity to the CCW (CW) of its nearest neighbor cavities. Only in the last non-symmetric cavity there is asymmetric coupling between the CW and CCW mode, i.e. CW is coupled to CCW with coupling strength $a$ but CCW is coupled to CW mode with coupling $b-\alpha$. This structure has a similar band structure as the one depicted in Fig.(\ref{fig2}). For example, for $a/t=1$ and $b/t=1.5$ we have an EP at $\alpha=1.5$ which is rebuts against disorder (not shown here).

{\it Conclusion}-- We have introduced the concept of robust EPs. We have started from an ordered system and find a recipe for obtaining the highest possible EP. We showed that disorder reduces the order of an EP. Using asymmetric tunneling, we show that one can obtain a topologically robust EP in a disordered system. The eigenstate associated with the robust EP is not localized, namely, it is extended. Our finding challenges the classical view related to the topological systems where a topological state is protected by some symmetries such as time reversal, chiral and particle-hole symmetries.


\begin{thebibliography}{0}


\bibitem{unitran2} Z. Lin, H. Ramezani, T. Eichelkraut, T. Kottos, H. Cao, and D.N. Christodoulides, Phys. Rev. Lett. {\bf 106}, 213901 (2011).
\bibitem{unitran3} S. Longhi, J. Phys. A  {\bf 44}, 485302 (2011).
\bibitem{unitran1} H. Ramezani, H.-K. Li, Y. Wang, and X. Zhang, Phys. Rev. Lett. {\bf 113}, 263905 (2014).
\bibitem{unitran4} Z. J. Wong, Y. L. Xu, J. Kim, K. O'Brien, Y. Wang, L. Feng, and X. Zhang, Nat. Photon. {\bf 10}, 796 (2016).
\bibitem{sense1} W. Chen, S. K. Ozdemir, G. Zhao, J. Wiersig, and L. Yang, Nature {\bf 548}, 192 (2017).
\bibitem{sense2} J. Wiersig, Phys. Rev. Lett. {\bf 112}, 203901 (2014).
\bibitem{listop} T. Goldzak, A. A. Mailybaev, and N. Moiseyev, Phys. Rev. Lett. {\bf 120}, 013901 (2018).



\bibitem{EP1} T. Kato, Perturbation Theory for Linear Operators (Springer-
Verlag, Berlin, 1966).
\bibitem{EP2} Ingrid Rotter, J. Phys. A: Math. Theor. {\bf 42}, 153001 (2009).
\bibitem{EP3} W. D. Heiss, J. Phys. A: Math. Theor. {\bf 45}, 444016 (2012). 


\bibitem{EPatom} Holger Cartarius, Jorg Main, and Gunter Wunner, Phys. Rev. Lett. {\bf 99}, 173003 (2007).
 
 

 
 
\bibitem{EPdeney1} C. Dembowski, H.-D. Graf, H. L. Harney, A. Heine, W. D. Heiss, H. Rehfeld, and A. Richter, Phys. Rev. Lett. {\bf 86}, 787 (2001).
\bibitem{EPdeney2} Jorg Doppler, et. al., Nature {\bf 537}, 76 (2016).
\bibitem{EPdeney3} Sang-Bum Lee, et. al., Phys. Rev. Lett. {\bf 103}, 134101 (2009).
\bibitem{EPdeney4} Kun Ding, Guancong Ma, Z.Q. Zhang, and C.T. Chan, Phys. Rev. Lett. {\bf 121}, 085702 (2018).


\bibitem{dynexcep1} Xu-Lin Zhang, Tianshu Jiang, Hong-Bo Sun, C. T. Chan, arXiv:1806.07649
\bibitem{dynexcep2} Absar U. Hassan, Bo Zhen, Marin Soljacic, Mercedeh Khajavikhan, and Demetrios N. Christodoulides, Phys. Rev. Lett. {\bf 118}, 093002 (2017).


\bibitem{EPmulti1} Jung-Wan Ryu, Soo-Young Lee, Sang Wook Kim, Phys. Rev. A {\bf 85}, 042101 (2012).
\bibitem{EPmulti2} Gilles Demange, Eva-Maria Graefe, J. Phys. A {\bf 45}, 025303 (2012) .
\bibitem{EPmulti3} Jing, S. K. Ozdemir, H. Lu and Franco Nori , Sci. Rep. {\bf 7}, 3386 (2017).


\bibitem{nonh2} C. Yuce, Phys. Lett. A {\bf 379} 1213 (2015).
\bibitem{PTop3} Kenta Esaki, Masatoshi Sato, Kazuki Hasebe, and Mahito Kohmoto, Phys. Rev. B {\bf 84} 205128 (2011).
\bibitem{ekl56} Julia M. Zeuner, et. al., Phys. Rev. Lett.  {\bf 115}, 040402 (2015).
\bibitem{PTop4} Pijush K. Ghosh, J. Phys.: Condens. Matter  {\bf 24}, 145302 (2012). 
\bibitem{sondeney1} S. Weiman, et. al., Nat. Mater. {\bf 16}, 433 ( 2017).
\bibitem{1d1} L. Jin, Phys. Rev. A {\bf 96}, 032103 (2017).
\bibitem{1d2} Chuanhao Yin, Hui Jiang, Linhu Li, Rong Lu, and Shu Chen, Phys. Rev. A {\bf 97}, 052115 (2018).
\bibitem{1d3} C. Yuce, Phys. Rev. A {\bf 98}, 012111 (2018).




\bibitem{2d1} Z Oztas, C Yuce, Phys. Rev. A  {\bf 98}, 042104 (2018).
\bibitem{2d3} Xue-Yi Zhu, et. al., arXiv:1801.10289. 
\bibitem{2d4} Kohei Kawabata, Ken Shiozaki, Masahito Ueda, arXiv:1805.09632.




\bibitem{bulkboun01} V. M. Martinez Alvarez, J. E. Barrios Vargas, L. E. F. Foa Torres, Phys. Rev. B {\bf 97}, 121401 (2018).
\bibitem{bulkboun02} Flore K. Kunst, Elisabet Edvardsson, Jan Carl Budich, Emil J. Bergholtz, Phys. Rev. Lett. {\bf 121}, 026808 (2018). 
\bibitem{bulkboun03} Shunyu Yao, Zhong Wang, Phys. Rev. Lett. {\bf 121}, 086803 (2018).
\bibitem{bulkboun04} Shunyu Yao, Fei Song, Zhong Wang, Phys. Rev. Lett.  {\bf 121}, 136802 (2018).
\bibitem{bulkboun05} C. Yuce, Phys. Rev. A {\bf 97}, 042118 (2018).
\bibitem{bulkboun06} Daniel Leykam, Konstantin Y. Bliokh, Chunli Huang, Y. D. Chong, and Franco Nori, Phys. Rev. Lett.  {\bf 118}, 040401 (2017).
\bibitem{bulkboun07} L. Jin, Z. Song, arXiv:1809.03139
\bibitem{winding1} Chuanhao Yin, Hui Jiang, Linhu Li, Rong Lu, and Shu Chen, Phys. Rev. A {\bf 97}, 052115 (2018).
\bibitem{winding2} Hui Jiang, Chao Yang, Shu Chen, arXiv:1809.00850.
\bibitem{winding3} Marcel Wagner, Felix Dangel, Holger Cartarius, Jorg Main, Gunter Wunner, Acta Polytechnica {\bf 57}, 470 (2017)


\bibitem{majo1} Yan Xing, Lu Qi, Ji Cao, Dong-Yang Wang, Cheng-Hua Bai, Wen-Xue Cui, Hong-Fu Wang, Ai-Dong Zhu, and Shou Zhang, Opt. Express {\bf 26}, 16250 (2018).
\bibitem{majo2} Marcel Klett, Holger Cartarius, Dennis Dast, Jorg Main, and Gunter Wunner, Phys. Rev. A {\bf 95}, 053626 (2017).
\bibitem{majo3} C. Yuce Phys. Rev. A {\bf 93}, 062130 (2016).
\bibitem{majo4} Henri Menke and Moritz M. Hirschmann, Phys. Rev. B {\bf 95}, 174506 (2017).
\bibitem{majo5} C. Li, X. Z. Zhang, G. Zhang and Z. Song, Phys. Rev. B {\bf 97}, 115436 (2018).
\bibitem{majo6} Ananya Ghatak, Tanmoy Das, Phys. Rev. B {\bf 97}, 014512 (2018).
\bibitem{nodalkun} Jan Carl Budich, Johan Carlstrom, Flore K. Kunst, Emil J. Bergholtz, arXiv:1810.00914.
\bibitem{nodalkun2} Zhesen Yang, Jiangping Hu, arXiv:1807.05661.
\bibitem{floquet1} C. Yuce, Eur. Phys. J. D {\bf 69}, 184 (2015).
\bibitem{floquet2} Z Turker, S Tombuloglu, C Yuce, Phys. Lett. A {\bf 382}, 2013 (2018). 
\bibitem{floquet3} Longwen Zhou, Qing-hai Wang, Hailong Wang, Jiangbin Gong Phys. Rev. A {\bf 98}, 022129 (2018).
\bibitem{floquet4} Longwen Zhou, Jiangbin Gong, arXiv:1807.00988.
\bibitem{lyang1} W. Chen, et al., Nature 548 (7666), 192 (2017)
\bibitem{mkhaje} H. Hodaei, et al., Nature 548 (7666), 187 (2017)
\bibitem{note} Although having $\beta=0$ makes the $\mathcal{H}$ non-invertible, Our choice $\beta\neq0$ does not change the results and it only shifts the real part of the energy level to zero.

\bibitem{52}B. E. Little, S. T. Chu, J. Haus, H. A. Foresi, and J.-P. Laine, J. Lightwave Technol. 15, 998 (1997).
\bibitem{53} N. Stefanou and A. Modinos, Phys. Rev. B 57, 12127 (1998).
\bibitem{54} A. Yariv, Y. Xu, R. K. Lee, and A. Scherer, Opt. Lett. 24, 711 (1999).
\bibitem{55} J. Wiersig, S. W. Kim, and M. Hentschel, Phys. Rev. A 78,053809 (2008).












\end{thebibliography}
\end{document}